\documentclass[11pt,twoside]{article}

\usepackage{asp2004}
\usepackage{epsf}
\usepackage{psfig}
\usepackage{lscape}

\begin{document}
\title{A Chemical Abundance Study of three RHB and two RGB stars in NGC~6637 (M69)} 

\author{Jae-Woo Lee}
\affil{Department of Astronomy and Space Science,
Astrophysical Research Center for the Structure and
Evolution of the Cosmos (ARCSEC\arcsec),
Sejong University, 98 Gunja-Dong, Gwangjin-Gu, Seoul, 143-747, Korea}

\author{Mercedes L\'opez-Morales}
\affil{Carnegie Institution of Washington, 
Department of Terrestrial Magnetism, 5241 Broad Branch Rd. NW, Washington DC 20015, USA}

\begin{abstract} 
We present a detailed chemical abundance study of 
three red horizontal branch and two red giant branch stars 
in the metal-rich globular cluster NGC 6637 (M69). 
The value of [Fe/H] derived from LTE calculations is $-$0.77 $\pm$ 0.02 dex. 
We also discuss the anticorrelation between oxygen and 
sodium abundances in the program stars and compare 
the [Si/Ti] ratio of NGC~6637 with those of other globular clusters.
\end{abstract}

\section{Introduction}
\label{sec:intro}

The Galactic bulge is one of the major component of our Galaxy,
however, its formation history is still not well understood.
There are two competing bulge formation scenarios. 
The first one is the merger hierarchical clustering
scenario, where bulges are built up during those mergers. 
In that scenario the dense central region of massive satellites 
may survive and sink to the center due to the tidal friction.
In this framework, only the most massive satellites 
could contribute to the central bulge formation in a Hubble time. 
The other scenario for bulge formation is an instability
in the disk. However, the Galactic bulge is dominated by old, 
metal-rich stars and neither of the two scenarios 
can explain this fact (Wyse \& Gilmore 2005).

The elemental abundances of globular clusters provide crucial information 
regarding the formation and the evolution of the Galaxy (Freeman \& Bland-Hawthorn 2002). 
Until recently, the number of globular clusters studied 
employing high-resolution spectroscopy was very limited 
due to the high interstellar extinction values 
towards the Galactic central region. 
Detailed chemical compositions of the Galactic bulge globular clusters 
has started to emerge during the last decade in the advent of large 
aperture telescopes with the high-resolution spectrographs.

We present in this work a detailed composition study of one of those clusters. 
NGC~6637 (M69) is an old, metal rich globular cluster approximately 
1.6 kpc away from the Galactic center. 
Heasley et al. (2000) studied NGC~6637 using the WFPC2 on  
the Hubble Space Telescope and concluded that 
it has an age similar to 47~Tuc (see also De Angeli et al. 2005). 
Previous metallicity measurements of NGC~6637 suggest that 
the metallicity of the cluster ranges from [Fe/H] $\approx$ $-$0.6 to $-$0.8. 
Zinn \& West (1984) derived [Fe/H] = $-$0.59 $\pm$ 0.19 from 
the color-magnitude diagram (CMD) morphology and the Q39 integrated light index. 
Later, Geisler (1986) obtained [Fe/H] = $-$0.6 from 
the Washington photometry system. 
More recently, Rutledge, Hesser, \& Stetson (1997 and references therein)  
measured the metallicity of the cluster using the Ca~II triplet lines 
of the RGB stars in near infrared passband and they obtained 
[Fe/H] = $-$0.72 $\pm$ 0.09 on the Zinn \& West's abundance scale and 
$-$0.78 $\pm$ 0.03 on the Carretta \& Gratton's abundance scale.

This study, which is tied directly to those of Lee \& Carney (2002) and 
Lee, Carney, \& Habgood (2005) and uses the same analysis methods, 
explores the detailed elemental abundances for two RGB stars and 
three red-horizontal branch (RHB) stars in NGC~6637.

\section{Observations and Data Reduction}
\label{sec:obs}

The observations were carried out during three different observing seasons. 
We selected our program stars from the $BV$ photometry of 
Sarajedini \& Norris (1994) and the 2MASS $JK$ photometry. 
In Table~1, we provide identification numbers (Hartwick \& Sandage 1968), 
$V$ magnitudes, $(B-V)$ colors (Sarajedini \& Norris 994) and  
$K$ magnitudes of our target stars.

The observations of the RGB stars were obtained using 
the CTIO 4-m telescope and its Cassegrain echelle spectrograph 
in July 1998 and June 1999. The Tek 2048 $\times$ 2048 CCD, 
31.6~lines/mm echelle grating, long red camera, and 
G181 cross-disperser were employed. The slit width was 150 $\mu$m, 
or about 1.0 arcsec. The projection of the slit on to 2.0 pixels yielded 
an effective resolving power $R$ = 28,000.
Each spectrum has complete spectral coverage from 5600 to 7800~\AA. 
All program star observations were accompanied by flat lamps, 
Th-Ar lamps, and bias frames. 
The raw data frames were trimmed, bias-corrected, and flat-fielded 
using the IRAF ARED and CCDRED packages. 
The scattered light was also subtracted using the APSCATTER task 
in the  ECHELLE package. The echelle apertures were then extracted 
to form 1-d spectra, which were continuum-fitted and normalized, 
and a wavelength solution was applied following 
the standard IRAF echelle reduction routines.

The observations of the RHB stars were obtained in July 2005 
with the Magellan Clay Telescope using the Magellan Inamori Kyocera  
Echelle spectrograph (MIKE; Bernstein et al.\ 2003). 
We used a 0.7 arcsec slit that provided a resolving power of 29,000 
in the red with wavelength coverage from 4950~\AA~to 7250~\AA. 
We used {\sc MIKE Redux} code\footnote{http://web.mit.edu/$\sim$burles/www/MIKE/}
to extract spectra which effectively correct for the tilted slit.

Equivalent widths were measured mainly by the direct integration 
of each line profile using the SPLOT task in IRAF ECHELLE package. 
We estimate the error in our measurement of the equivalent width 
to be $\pm$ 2 -- 3 m\AA. The main sources of error are noise features 
in the spectra and our ability to determine the proper continuum level.

\begin{table}[!t]
\caption{Program stars in NGC~6637}
\smallskip
\begin{center}
{\small
\begin{tabular}{lccccccc}
\tableline
\noalign{\smallskip}
ID & $V$ & $(B-V)$ & $K$  & $T_{eff}$ & $\log g$ & $v_{turb}$ & [Fe/H] \\
   &     &         &      & (K)       &          & (km/s) & \\
\noalign{\smallskip}
\tableline
\noalign{\smallskip}
RGB &   &   &   &   &  &  & \\
I-30 & 13.59 & 1.68 & ~9.63   & 3890 & 0.8 & 1.85 & $-$0.78 \\
I-6  & 13.86 & 1.70 & ~9.53   & 3900 & 0.7 & 1.85 & $-$0.76 \\
  &   &   &   &   &  &  & \\
RHB &   &   &   &   &  &  & \\
I-37 & 15.85 & 0.92 & 13.52   & 5175 & 2.5 & 2.10 & $-$0.79 \\
III-37 & 15.95 & 0.93 & 13.50 & 5025 & 2.5 & 1.80 & $-$0.73 \\
IV-1 & 15.96 & 0.86 & 13.57   & 5100 & 2.5 & 1.46 & $-$0.78 \\

\noalign{\smallskip}
\tableline\
\end{tabular}
}
\end{center}
\end{table}

\section{Analysis}
\label{sec:analysis}

For our line selection, laboratory oscillator strengths 
were adopted whenever possible, with supplemental solar oscillator 
strength values. In addition to oscillator strengths, 
taking into account the damping broadening due to the van der Waals force, 
we adopted the Uns\"old approximation with no enhancement 
(Lee \& Carney 2002; Lee, Carney, \& Habgood 2005).

For the analysis, we rely on spectroscopic temperatures and 
photometric surface gravities, following the method described in 
Lee \& Carney (2002). 
The initial estimates of the temperature of our program stars
were calculated using their $BVK$ photometry and 
the empirical color-temperature relations given by 
Alonso, Arribas, \& Martinez-Roger (1999). 
To derive the stars' photometric surface gravity in relation 
to that of the Sun, we used $\log g_{\sun}$ = 4.44 in cgs units, 
$M_{\rm bol,\sun}$ = 4.74 mag, and $T_{\rm eff,\sun}$ = 5777 K.

The abundance analysis was performed using the current version
of the local thermodynamic equilibrium (LTE) line analysis
program MOOG (Sneden 1973).
For input model atmospheres, we interpolated Kurucz models
using a program kindly supplied by A.\ McWilliam (2005, private communication).
Adopting the photometric temperature and surface gravity
as our initial values, we began by restricting the analysis to
those \ion{Fe}{1} lines with $\log$(W$_{\lambda}$/$\lambda$) $\leq$ $-5.2$
(i.e., for  the linear part of the curve of growth),
and comparing the abundances as a function of excitation potential.
New model atmospheres were computed with a slightly different effective
temperature until the slope of the $\log$~n(\ion{Fe}{1}) 
versus excitation potential relation was zero to within the uncertainties.
The stronger \ion{Fe}{1} lines were then added and the microturbulent
velocity $v_{\rm turb}$ altered until the $\log$ n(\ion{Fe}{1}) versus
$\log$(W$_{\lambda}$/$\lambda$) relation had no discernible slope.
Table~1 shows temperatures and surface gravities 
for our program stars.
 
\begin{figure}[!h]
\plotone{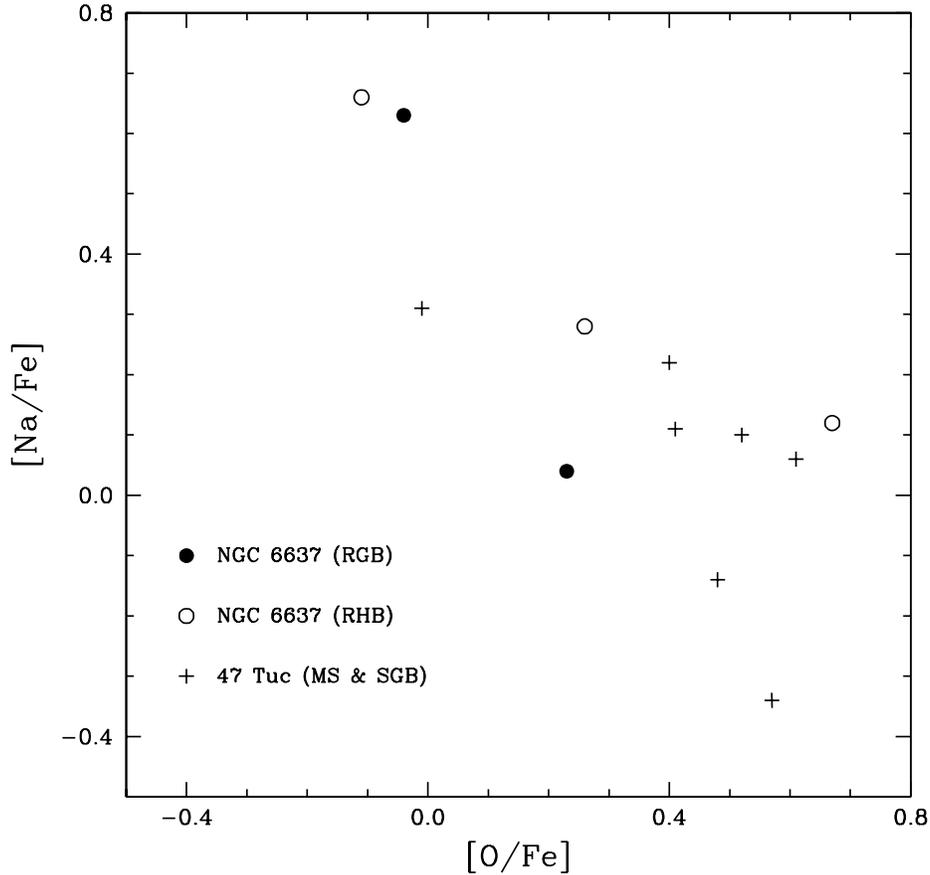}
\caption{A comparison of [Na/Fe] vs.\ [O/Fe] for MS and SGB stars in
 47~Tuc, and RGB and RHB stars in NGC~6637.}
  \label{fig:NaO}
\end{figure}

\begin{figure}[!t]
  \plotone{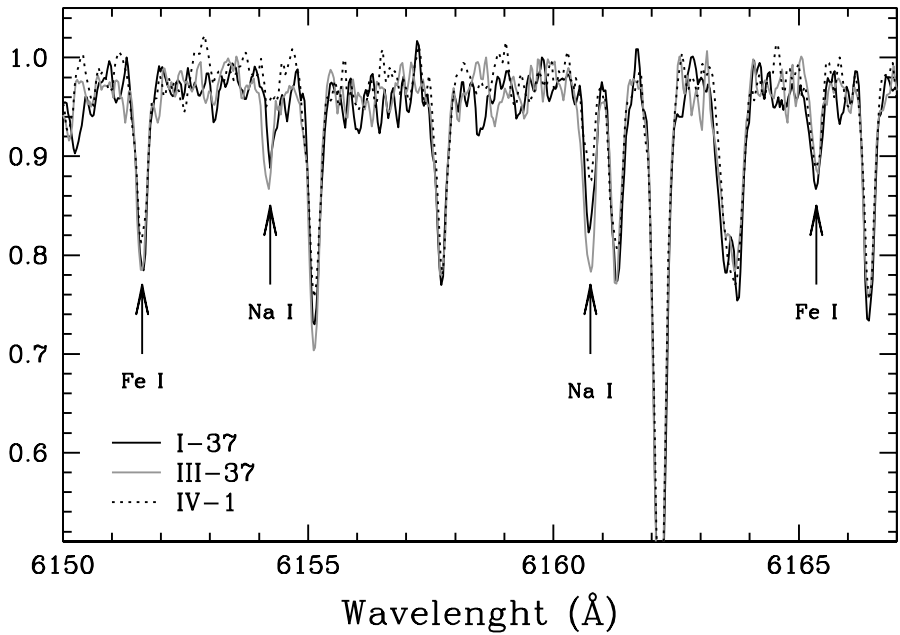}
  \caption{A comparison of observed spectra of three RHB stars in NGC~6637.}
  \label{fig:spectra}
\end{figure}

\begin{figure}[!t]
  \plotone{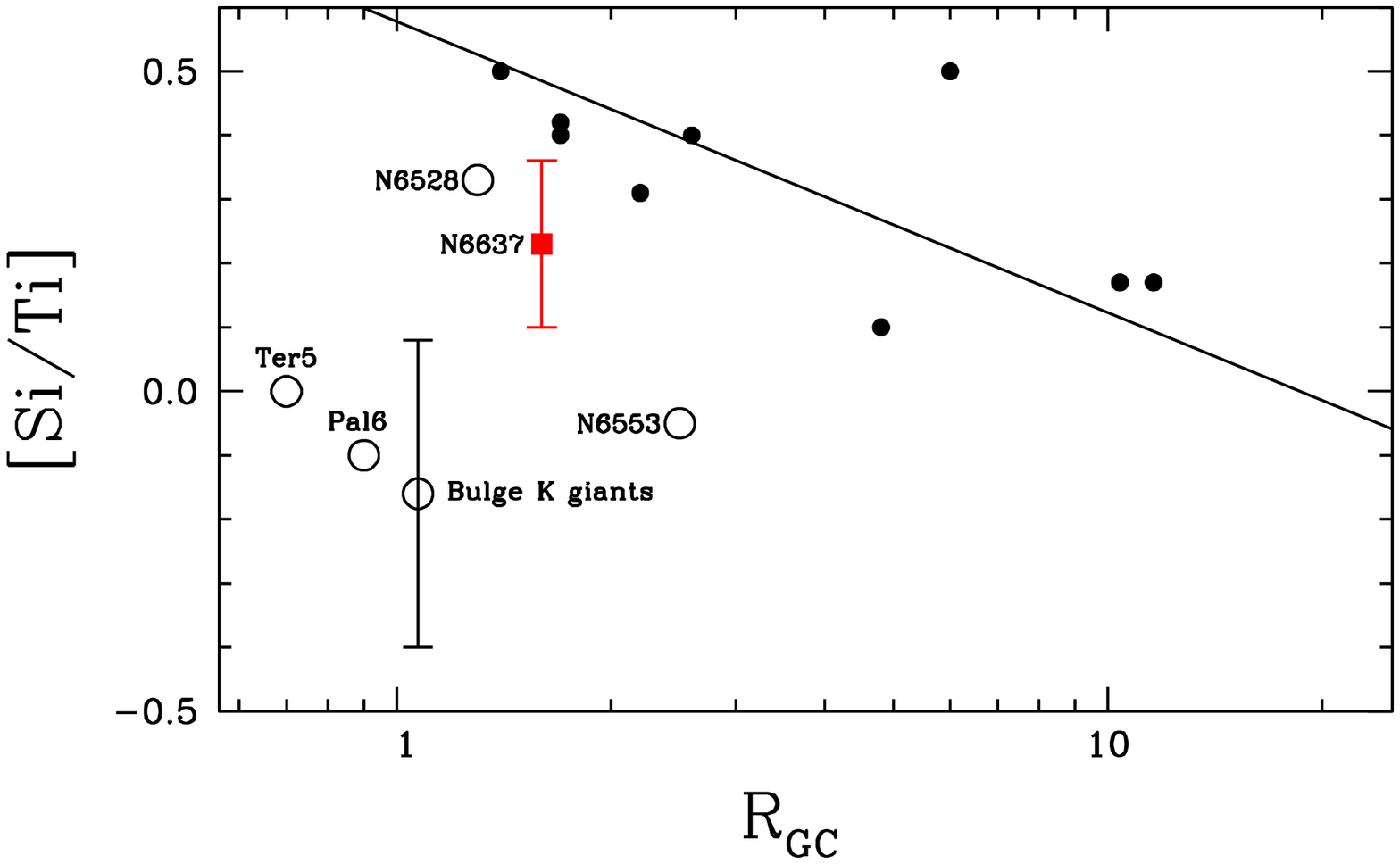}
  \caption{[Si/Ti] as a function of $R_{GC}$.
Dots represent 9 OHGCs analyzed by Lee (2006 in prep.).
The solid line represents the bisector linear fit to the data.
A non-parametric Spearman rank-order test
indicates a probability of $\approx$ 0.09~\% that 
the anti-correlation between [Si/Ti] and $R_{GC}$ of the clusters is random.}
  \label{fig:siti}
\end{figure}

\section{Results and Discussion}
\subsection{A Na-O anticorrelation in RHB stars in NGC~6637: 
Mixing or Primordial Variations? }

Many globular clusters appear to show anticorrelations 
between the abundances of oxygen and sodium, 
and of magnesium and aluminum. 
The subject was reviewed by Kraft (1994), and has been revisited 
by numerous authors. 
The approach often taken has been that these anticorrelations 
arise from deep mixing, whereby material whose chemical composition 
has been altered by proton captures within the CNO cycle is brought 
to the stellar surface. 
This concept has become less plausible with the discovery 
that such anticorrelations are also seen in relatively unevolved stars 
in the metal-poor clusters NGC~6397 and NGC~6752 (Gratton et al.\ 2001) 
and in the metal-rich cluster 47~Tuc (Carretta et al.\ 2004).

In Figure~\ref{fig:NaO} we show a plot of [Na/Fe] vs.\ [O/Fe] for NGC~6637 
in comparison with the results for main sequence stars (MS) and 
sub-giant stars (SGB) in 47~Tuc from Carretta et al.\ (2004). 
The figure reveals that an anticorrelation between [Na/Fe] and [O/Fe] 
appears to exist not only in RGB stars but also in RHB stars in NGC~6637. 
In Figure~\ref{fig:spectra}, we show a comparison  of observed spectra 
of three RHB stars in NGC~6637. 
As shown in Table~1, these stars have almost identical stellar physical parameters. 
Figure~\ref{fig:spectra} shows that absorption line strengths 
of two iron lines at $\lambda$6151.62\AA\ and $\lambda$6165.36\AA\ 
are identical, while those of sodium lines at $\lambda$6154.23\AA\ and 
$\lambda$6160.75\AA\ vary significantly among RHB stars, 
indicating that the sodium abundance variations are real. 
This [Na/Fe] vs.\ [O/Fe] anticorrelation in RHB stars in NGC~6637 cannot be
understood by the deep mixing scenario.

\subsection{[Si/Ti] ratios of globular clusters}
A recent study by Lee \& Carney (2002) found that the [Si/Ti] ratio 
in old halo globular clusters (OHGC) increases towards the Galactic center. 
To interpret this finding they propose that 
(i) the inner parts of the Galaxy have been metal-enriched 
by Type II supernovae (SNe~II) explosions of stars that are 
more massive than the stars in the outer parts, and 
(ii) the metal enrichment in the inner parts of the Galaxy is 
further enhanced by the higher density of material in those regions, 
which cause a higher retention rate of the material expelled by the supernovae.

In Figure~\ref{fig:siti} we show [Si/Ti] of NGC~6637 and
9 OHGC (solid circles) as a function of Galactocentric distance.
It should be emphasized that these 9 OHGC were observed with the
same instrument setups (the CTIO 4m telescope with its echelle
spectrograph) and were analyzed employing the same method
(Lee 2006 in prep.).
In the figure we show the fit to the data. A non-parametric
Spearman rank-order test gives a probability of $\approx$ 0.09\%
that the anti-correlation between [Si/Ti] ratios and Galactocentric distances 
of OHGCs is random.
Also illustrated in the figure are 3 metal-rich globular clusters
NGC~6528 (Carretta et al.\ 2001), NGC~6553 (Cohen et al.\ 1999),
and Palomar~6 (Lee et al.\ 2004).

NGC~6637 and NGC~6528 seem to follow [Si/Ti] abundance ratios
found by Lee \& Carney (2002). NGC~6553 and Palomar~6, on the other hand,
seem to have [Si/Ti] abundances similar to galactic bulge K giants
(McWilliam \& Rich 1994), 
where the Ti abundances are enhanced.
There seem to be therefore,
two different groups of metal-rich globular clusters in the inner Galaxy,
one that follows the [Si/Ti] trend of old halo stars,
and another group that has [Si/Ti] ratios similar to the giants stars
in the Bulge, indicating that these two different groups were
formed in different physical environments.

\acknowledgements
This research was supported by the Korea Science
and Engineering Foundation (KOSEF) to the Astrophysical Research Center
for the Structure and Evolution of the Cosmos (ARCSEC)
by the Carnegie Institution of Washington through a Carnegie fellowship.

\end{document}